\documentclass{PoS}

\usepackage{cite}
\usepackage{amsmath, amsthm, amssymb}
\usepackage{url}
\usepackage{graphicx}

\newcommand{\HA}{\text{H}}
\newcommand{\GeV}{\text{GeV}}

\title{{\footnotesize DESY 19--186, DO-TH 19/25, SAGEX-19-26,
       PoS(RADCOR19)046}\\
Revisiting the ${O(\alpha^2)}$ Initial State QED Corrections to $e^+ \, e^-$ 
Annihilation 
into a Neutral Boson}

\ShortTitle{$O(\alpha^2)$ Initial State Corrections}

\author{\speaker{Johannes Bl{\"u}mlein} \\
      Deutsches Elektronen--Synchrotron, DESY,
      Platanenallee 6, D-15738 Zeuthen, Germany.\\
      E-mail: \email{johannes.bluemlein@desy.de}
}

\author{Abilio De Freitas \\
      Deutsches Elektronen--Synchrotron, DESY,
      Platanenallee 6, D-15738 Zeuthen, Germany.\\
      E-mail: \email{abilio.de.freitas@desy.de}
}

\author{Clemens Raab \\
      Johannes Kepler Universit{\"a}t Linz,
      Altenbergerstra\ss{}e 69, A-4040 Linz, Austria.\\
      E-mail: \email{clemens.raab@jku.at}
}

\author{Kay Sch{\"o}nwald \\
      Deutsches Elektronen--Synchrotron, DESY,
      Platanenallee 6, D-15738 Zeuthen, Germany.\\
      E-mail: \email{kay.schoenwald@desy.de}
}

\abstract{
At $e^+ \, e^-$ colliders the QED--initial state radiation forms a large part of the radiative corrections. 
Their precise and fast evaluation is an essential asset for the experiments at LEP, the ILC and the FCC-ee,
operating at high luminosity. A long standing problem in the analytic calculation of the $O(\alpha^2)$ 
initial state corrections concerns a discrepancy which has been observed between the result of Berends 
et al. (1988) \cite{Berends:1987ab} in the limit $m_e^2 \ll s$ and the result by Bl{\"u}mlein et al. 
(2011) \cite{Blumlein:2011mi} using massive operator matrix elements deriving this limit directly. 
In order to resolve this important issue we recalculated this process by integrating directly over the 
phase space without any approximation. For parts of the corrections we find exact solutions of the cross 
section in terms of iterated integrals over square root valued letters representing incomplete elliptic 
integrals and iterations over them. The expansion in the limit $m_e^2 \ll s$ reveals errors in the 
constant $O(\alpha^2)$ term of the former calculation and yields agreement with the calculation based on 
massive operator matrix elements, which has impact on the experimental analysis programs. This finding also
explicitly proofs the factorization of massive initial state particles in the high energy limit 
including the terms of $O(\alpha^2)$ for this process.}

\FullConference{14th International Symposium on Radiative Corrections (RADCOR2019)\\ 
9-13 September 2019\\
		Palais des Papes, Avignon, France}

\begin{document}
\section{Introduction} 
\label{sec:1}

\vspace*{1mm}
\noindent
At $e^+ \, e^-$ colliders a major part of the radiative corrections is due to QED initial state 
radiation (ISR). Therefore the precise knowledge of these corrections is of fundamental importance 
for the precision measurements at LEP \cite{ALEPH:2005ab} and future colliders like the ILC and CLIC 
\cite{ILC}, FCC-ee \cite{Abada:2019zxq,FCCEE} and CEPC \cite{CEPC}. A first calculation of the 
$O(\alpha^2)$ corrections to the process $e^+ \, e^- \to \gamma^*/Z^*$ has already been performed in 
1987 \cite{Berends:1987ab} neglecting power corrections in the electron mass. These results have been 
used in the analysis of the LEP1 data in various fitting codes like \texttt{TOPAZ0} \cite{Montagna:1998kp} 
and \texttt{ZFITTER} \cite{ZFITTER} at the $Z$ peak. These corrections will be also important 
for future measurements of the $t \, \bar{t}$ production and associated Higgs production in the process
$e^+ \, e^- \to Z^* \, H$.

A second calculation of the $O(\alpha^2)$ QED ISR corrections has been performed in 2011 
\cite{Blumlein:2011mi} by using the light cone expansion and assuming the factorization of the massive 
Drell-Yan process.
The differential cross section factorizes in this process 
\begin{eqnarray}
      \frac{d \sigma_{e^+ \, e^-}}{d s^\prime} &=&
      \frac{1}{s} \sigma_{e^+ e^-}^{(0)}(s^\prime) H \left(z,\alpha,\frac{s}{m^2_e}\right)
\end{eqnarray}
into the Born cross section $\sigma_{e^+ e^-}^{(0)}$ and the radiator function $H(z,\alpha,\frac{s}{m^2})$.
Here $s$ is the center-of-mass energy of the process, $s^\prime$ the invariant
mass of the off-shell vector boson, $z=s^\prime/s$ and $\alpha=\alpha(s)$ is the fine structure constant.
The radiator functions can be expanded in $\alpha$ 
\begin{eqnarray}
      H(z,\alpha,\frac{s}{m^2}) &=& \delta(1-z) 
      + \sum\limits_{k=1}^{\infty} \left( \frac{\alpha}{4\pi} \right)^k C_k \left(z,\frac{s}{m_e^2} \right) .
\end{eqnarray}
In the limit $m^2_e \ll s$, dropping power corrections in $m_e^2/s$, the 
functions $C_k$ can be expressed via
\begin{eqnarray}
      C_k \left(z,\frac{s}{m_e^2} \right) &=& \sum\limits_{l=0}^{k}
      \ln^{k-l} \left( \frac{s}{m^2_e} \right) c_{k,l}(z) .
\end{eqnarray}
The two calculations show differences in the non-logarithmic terms at
$O(\alpha^2)$, i.e. the coefficient $c_{2,2}(z)$ does not agree.

To clarify this discrepancy we have repeated the calculation by following the steps of Berends et al. 
but without expanding in the small ratio $m_e^2/s$ in intermediate steps. For many processes we find 
solutions in terms of iterated integrals over square-root valued letters, which can be expanded in the
limit $m_e^2 \ll s$. In Section~\ref{sec:2}~we will give a short review of our method of calculation. 
The results of these calculations are discussed in Section~\ref{sec:3}~and the differences between the 
two calculations are quantified. In Section~\ref{sec:4}~we show the phenomenological importance of 
these corrections on the $Z$-boson peak and $t \, \bar{t}$-production.
Section~\ref{sec:5}~contains the conclusions.  
\section{Calculation} 
\label{sec:2}

\vspace*{1mm}
\noindent
At $\mathcal{O}(\alpha^2)$ the radiators due to the real emission of an 
additional fermion pair or of two photons from the initial state can be calculated
via the integration over the $2 \to 3$ phase space integral over the corresponding
matrix element. The phase space can then be parameterized as a four-fold integral
over two angles and two kinematic invariants.
Since the integration over the angles introduces at most one square-root,
the integration over the first invariant can be performed using standard
techniques after rationalization of the appearing square-root.
After this integration the last integrand is given in terms of a 
linear combination of rational functions, logarithmic and dilogarithmic
expression containing multiple square-roots which cannot be rationalized
simultaneously.
To deal with the last integral we use the algorithm described in more 
detail in \cite{Blumlein:2019qze,Blumlein:2019zux} to express the result in terms of iterated 
integrals over square root valued letters. The code is written in \texttt{Mathematica} and uses the 
routine  \texttt{DSolveRational} of the package \texttt{HolonomicFunctions} \cite{KOUTSCHAN}.
For the general theory see \cite{RaabRegensburger,Raab} and \cite{GuoRegensburgerRosenkranz}, 
where the simpler case without endpoint singularities is discussed.
The main steps can be summarized as follows:
First, the logarithms and dilogarithms in the integrand are rewritten 
in terms of nested integrals evaluated at the last integration variable using first-order differential 
equations. This introduces a first set of letters needed to express the final result.
Then the prefactors are now rewritten in terms of this set of letters
and if this is not possible new letters are introduced into the alphabet.
Finally, the last integration can be done iteratively.
Since the phase space is more complicated compared to the deep-inelastic
kinematics discussed in \cite{Blumlein:2019qze,Blumlein:2019zux} we find
a more involved set of letters.
In total we find a set of 37 new letters, the most involved ones are given
by
\begin{eqnarray}
      f_{w_1}(t) &=& \frac{1}{\sqrt{1-t}\sqrt{t^2(1-z)^2-8\rho(1+z)t+16\rho^2}} ,
\\
      f_{w_2}(t) &=& \frac{1}{\sqrt{t}\sqrt{1-t}\sqrt{t^2(1-z)^2-8\rho(1+z)t+16\rho^2}} ,
\\
      f_{w_3}(t) &=& \frac{\sqrt{t}}{\sqrt{1-t}\sqrt{t^2(1-z)^2-8\rho(1+z)t+16\rho^2}} ,
\end{eqnarray}
where we used $\rho=m_e^2/s$ and $z=s^\prime/s$ with $s^\prime$ being the invariant mass of the off-shell 
vector boson. Subsequently the iterated integrals can be expanded in the asymptotic limit $m_e^2 \ll 
s$. Although the expansion of single iterated integrals gives still rise to iterated integrals over 
(much simpler) square root valued letters, the full expressions in the asymptotic limit can be 
expressed by harmonic polylogarithms.
\section{Results} 
\label{sec:3}

\begin{figure}
      \centering
      \includegraphics[width=0.75\textwidth]{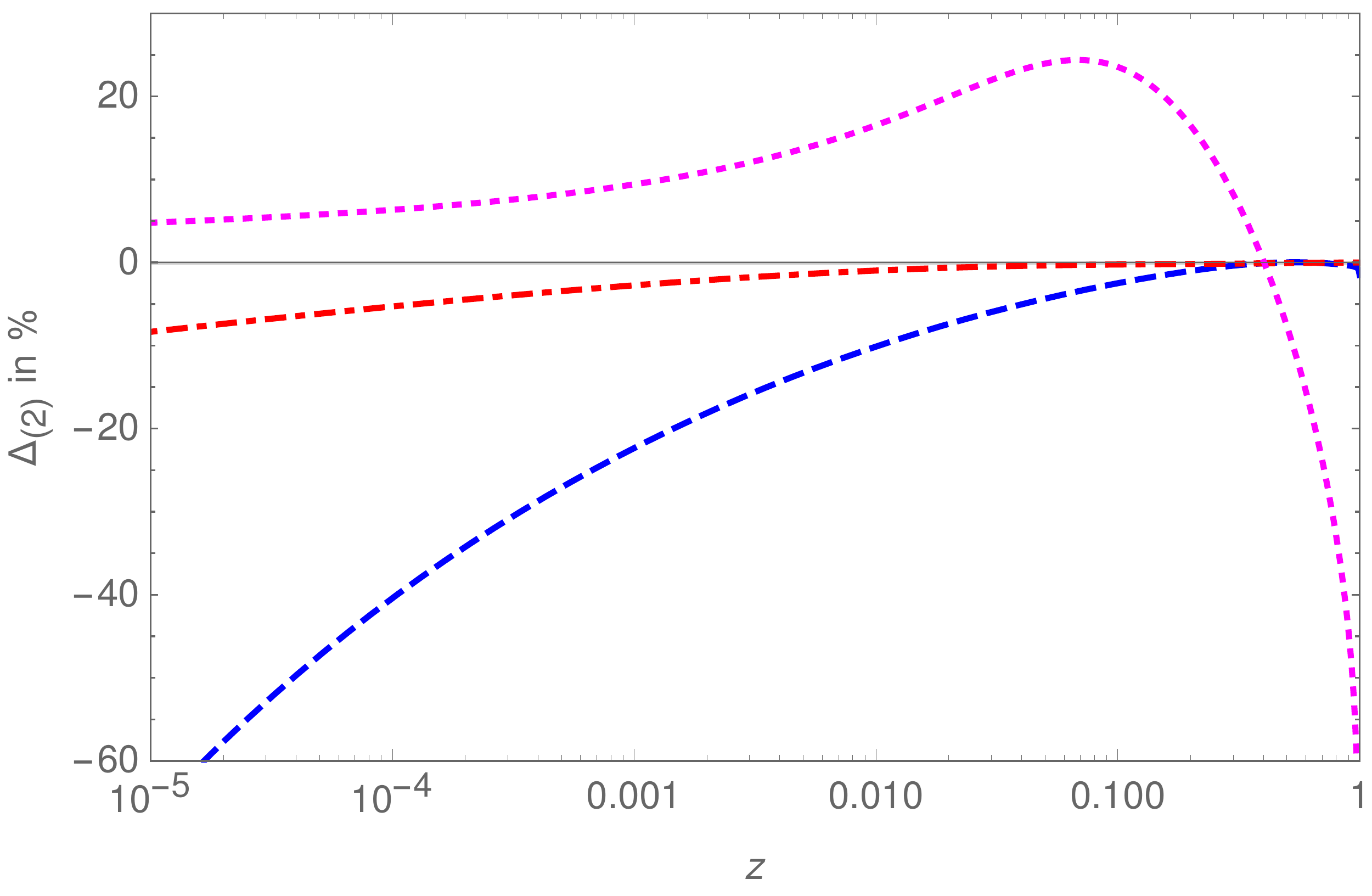}
      \caption{Relative deviations of the results of Ref.~\cite{Berends:1987ab}~from the exact result in \% for the $O(\alpha^2)$ 
  corrections. 
  The non--singlet contribution (process II): dash-dotted line;
  the pure singlet contribution (process III): dashed; 
  the interference term between both contributions (process IV): dots; for $s = M_Z^2$, $M_Z = 
91.1879$~GeV; from \cite{Blumlein:2019srk}.}
      \label{fig:RAT}
\end{figure}

\vspace*{1mm}
\noindent
After expanding our results for the radiators in the asymptotic limit we find differences compared to the 
results obtained in \cite{Berends:1987ab} in the non-logarithmic terms. Following the notation 
in \cite{Berends:1987ab} we will refer to the different contributions to the radiator as processes I to 
IV. Process I describes the contributions from photon radiation, process II denotes the non-singlet and 
process III pure singlet fermion pair emission, while process IV results from the interference of the 
diagrams of the latter two processes. We use the quantities
\begin{eqnarray}
      \delta_i (z) &=& C_{2}^{i}(z) - C_{2}^{i,\text{BBN}}(z) ,
\\
      \Delta_i (z) &=& \frac{\delta_i (z)}{C_{i}(z)}, \qquad \qquad \qquad i=\text{I, II, III, IV},
\end{eqnarray}
where $C_{2}^{i,\text{BBN}}(z)$ stand for the radiators of Ref~\cite{Berends:1987ab}~to quantify the 
discrepancies between the calculations. The results for the fermion pair emission, i.e. processes II to 
IV, have already been given in \cite{Blumlein:2019srk}. The calculation of process I has been finished 
recently and we find
\begin{eqnarray}
      \delta_\text{I} (z) &=&  - 8 + \frac{100}{33} z \HA_0^2 
      + 8 \HA_1 + 4(1-z)\HA_1^2 - \frac{8(2-z)z}{1-z} \HA_{0,1} + \frac{8(2-2z+z^2)}{1-z} \zeta_2  .
\end{eqnarray}
The sum of processes I and IV yields what has been called process I in Ref.~\cite{Blumlein:2011mi}, 
since the method of operators matrix elements cannot distinguish between these processes. The sum of the 
newly calculated process I and process IV agrees with the result obtained in Ref.~\cite{Blumlein:2011mi}~and 
proofs thereby the factorization of the massive Drell-Yan process in the asymptotic limit to 
$O(\alpha^2)$. The quantity $\Delta_i (z)$ measures the discrepancy of the radiators and is plotted in 
Figure~\ref{fig:RAT}. We see a $20-60\%$ discrepancy between the $O(\alpha^2)$ radiators, with the 
largest deviation arising from process III at low values of $z$.  

\begin{figure}
\centering
\begin{minipage}{\textwidth}
\includegraphics[width=\textwidth]{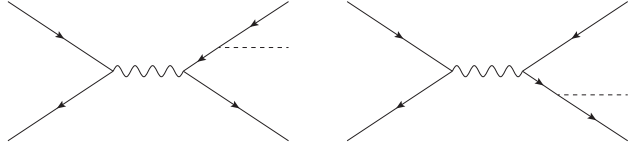}
\end{minipage}

\vspace*{-16cm}
\caption{Graphs not considered in the calculation of the initial state $O(\alpha^2)$ corrections to $\gamma^*/Z^*$ production due to $e^+ e^-$ pair production in Ref.~\cite{Berends:1987ab}.}
\label{fig:ISR-not}
\end{figure}

\begin{figure}
      \centering
      \includegraphics[width=0.75\textwidth]{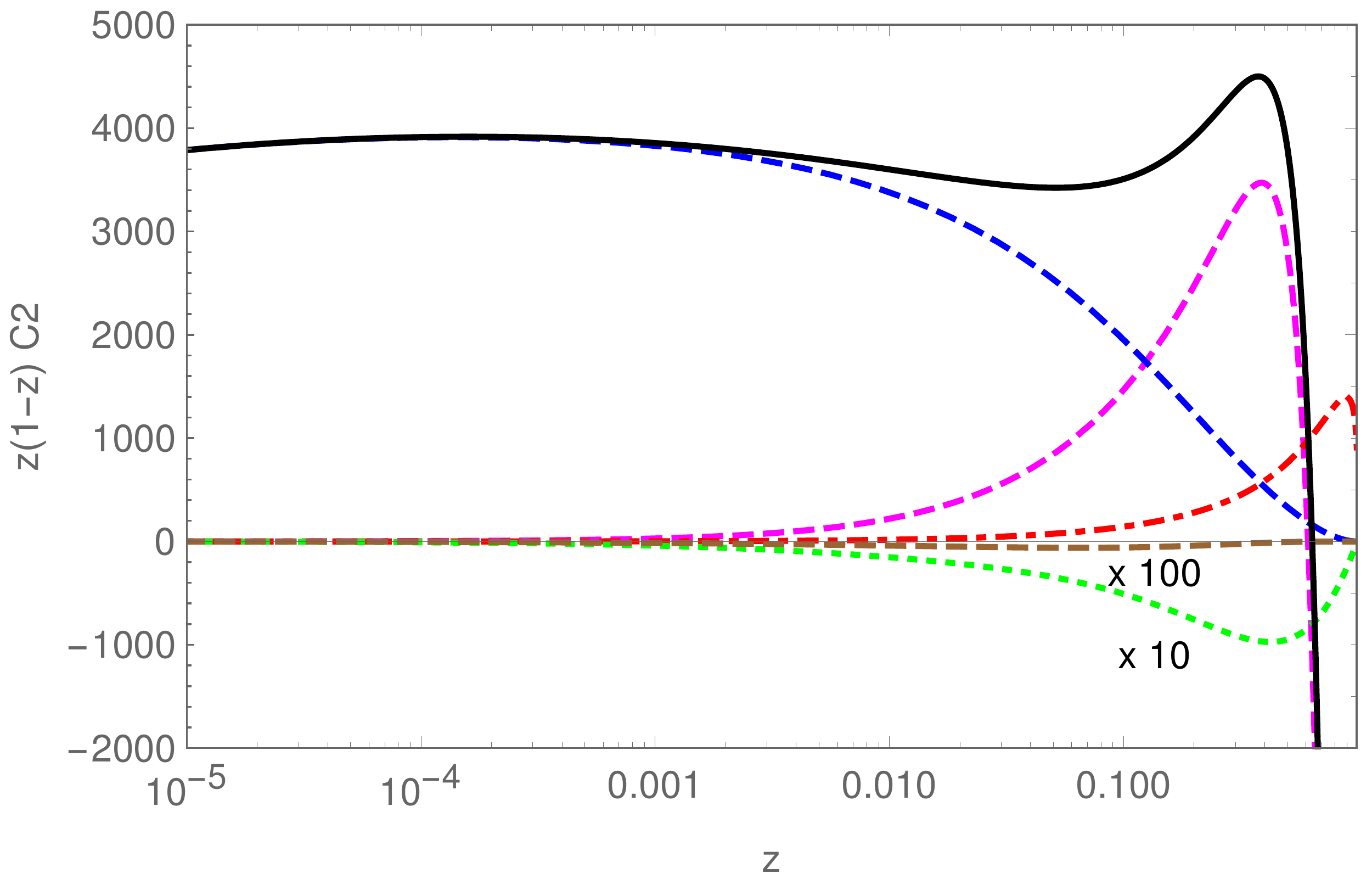}
      \caption{The initial state $O(\alpha^2)$ corrections to $\gamma^*/Z^*$ production due to $e^+ e^-$ pair production multiplied by 
$z(1-z)$. 
  The photon contribution (process I): dashed, magenta line;
  The non--singlet contribution (process II): dash-dotted, red line;
  the pure singlet contribution (process III): dotted, green line; 
  the interference term between both contributions 
  (process IV) $\times 10$: dashed, blue; the vector contributions implied by process B, Ref.~\cite{Hamberg:1990np}, and interferences (process V) 
$\times 100$: 
dashed, brown line; all contributions: full, black line for $s = M_Z^2$; from \cite{DYLONG}.}
      \label{fig:C2}
\end{figure}

Additionally to the discrepancies above there are contributions not accounted for in 
Ref.~\cite{Berends:1987ab}. They arise from the graphs given in Figure~\ref{fig:ISR-not}~and their 
interference with the non-singlet and pure-singlet diagrams. In the following we will summarize these 
contributions as process V. These contributions do not contain logarithms in $\rho$ and after expanding
we recover the results of the massless calculation in Ref.~\cite{Hamberg:1990np}. The total values of 
the different radiators can be seen in Figure~\ref{fig:C2}. One sees that process I dominates at $z \to 1$, 
where also process II is not negligible, while process III dominates for small values of $z$. The other 
radiators are numerically small. 
\section{Phenomenology} 
\label{sec:4}

\vspace*{1mm}
\noindent
In the  following we want to present some phenomenological studies of the $O(\alpha^2)$ initial state radiation.
First we will discuss the inclusive cross section of $\mu^+ \, \mu^-$ production 
before turning to the $t \, \bar{t}$ production at threshold.
For more details and information on $t \, \bar{t}$ production in the continuum and associated Higgs boson production 
see Ref.~\cite{Blumlein:2019pqb}.

\subsection{The $Z$-peak} 
\label{sec:41}

\begin{figure}
  \centering
  \hskip-0.8cm
  \includegraphics[width=.7\linewidth]{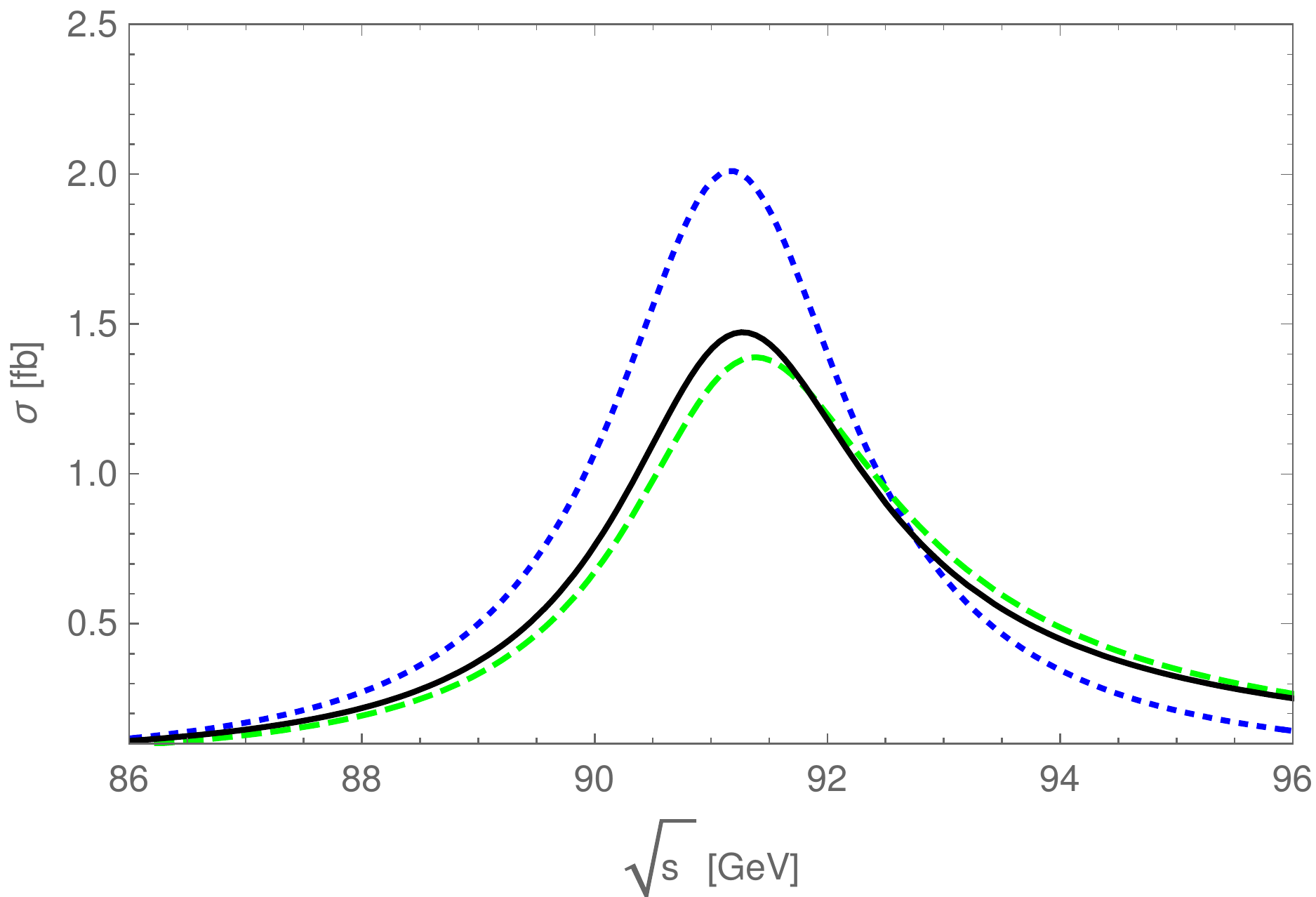}
  \caption{The $Z$-resonance in $e^+e^- \rightarrow \mu^+\mu^-$. Dotted line: Born cross section; Dashed line: 
  $O(\alpha)$ ISR corrections; 
  Full line: $O(\alpha^2)$ + soft resummation ISR corrections, with $s_0 = 4 m_\tau^2$; from 
\cite{Blumlein:2019pqb}.} 
  \label{fig:ZPEAK2}
\end{figure}

\vspace*{1mm}
\noindent
In Figure~\ref{fig:ZPEAK2}~the cross section of the production of a $\mu^+ \, \mu^-$ pair above the threshold
$s_0 = 4 m_\tau^2$, with $m_\tau$ the mass of the tau lepton, of its invariant mass around the Z peak with different orders of the initial state radiation
accounted is shown. 
The lines for the $O(\alpha^2)$ correction and the one including soft resummation are nearly identical.
The theoretical threshold is $s_0 = 4 m_\mu^2$, while in measurements
different cuts and a later extrapolation to the physical value was used.
In analysis at LEP1 examples for the experimental cuts are
$s_0 = 4 m_\tau^2$ or $s_0= 0.01 M_Z^2$~\cite{ALEPH:2005ab}.
In the following we will restrict the discussion to the first cut.
In general the discrepancies with the results obtained in Ref.~\cite{Berends:1987ab}
become larger for smaller values of $s_0$ since terms $\propto 1/z$ which
arise in process III were not correctly obtained there.

In Table~\ref{TAB1}~the shifts in the position and half-width of the $Z$ peak
of the different order ISR corrections performing the difference from one to the previous order are shown. 
Very similar values are obtained comparing the use of a fixed width or the $s$-dependent width. 
The peak shifts by $34.2$ MeV and the width width by 1 MeV regardless of the applied ISR corrections, in accordance with Refs.~\cite{SDEP}.

Including the $e^+e^-$ pair production to the pure photon emission at
$O(\alpha^2)$ shifts the width by 28 MeV while the peak position is unchanged. 
The exponentiation of soft photons from $O(\alpha^3)$ onward shifts the peak by 17 MeV and the width by 23 MeV.
At $s_0 = 4 m_\tau^2$ the differences to the radiators of Ref.~\cite{Berends:1987ab} are not resolvable using the peak position. 
However, the width is shifted by 4 MeV compared to the present result.
Since the current error of the half-width of the $Z$ boson 
is $\Delta \Gamma_Z = \pm 2.3$ MeV \cite{PDG} this of relevance 
even for LEP1 data.
The shift of the peak position of 0.2 MeV is relevant at  Giga-Z and Fcc\_ee \cite{ILC,Abada:2019zxq}, where resolutions of a few 
hundred  keV can be reached  for both $M_Z$ and $\Gamma_Z$, see also \cite{dEnterria:2016sca}.
\begin{table}
\centering
\begin{tabular}{|l|r|r|r|r|}
\hline
\multicolumn{1}{|c|}{} &
\multicolumn{2}{c|}{Fixed width} &
\multicolumn{2}{c|}{$s$ dep. width} \\
\hline
\multicolumn{1}{|c|}{} &
\multicolumn{1}{c|}{Peak} &
\multicolumn{1}{c|}{Width}    &
\multicolumn{1}{c|}{Peak} &
\multicolumn{1}{c|}{Width} \\
\multicolumn{1}{|c|}{} &
\multicolumn{1}{c|}{(MeV)} &
\multicolumn{1}{c|}{(MeV)}    &
\multicolumn{1}{c|}{(MeV} &
\multicolumn{1}{c|}{(MeV)} \\
\hline 
$O(\alpha)$   correction                 &  210 &  603 &  210 &  602 \\
$O(\alpha^2)$ correction                 & -109 & -187 & -109 & -187 \\
$O(\alpha^2)$: $\gamma$ only             & -110 & -215 & -110 & -215 \\
$O(\alpha^2)$ correction                 &      &      &      &      \\
+ soft exp.                              &   17 &   23 &   17 &   23 \\
Difference to $O(\alpha^2)$ \cite{Berends:1987ab}      &      &    4 &      &    4 \\
\hline
\end{tabular}
\caption{Shifts in the $Z$-mass and the width due to the different contributions to the ISR QED 
radiative corrections for a fixed width of $\Gamma_Z =  2.4952~\GeV$  and $s$-dependent width 
using $M_Z = 91.1876~\GeV$  \cite{PDG} and $s_0 = 4 m_\tau^2$, cf.~\cite{ALEPH:2005ab}; from 
\cite{Blumlein:2019pqb}.}
\label{TAB1}
\end{table}

\subsection{$t \, \bar{t}$ production}
\label{sec:42}

\begin{figure}
  \centering
  \hskip-0.8cm
  \includegraphics[width=.7\linewidth]{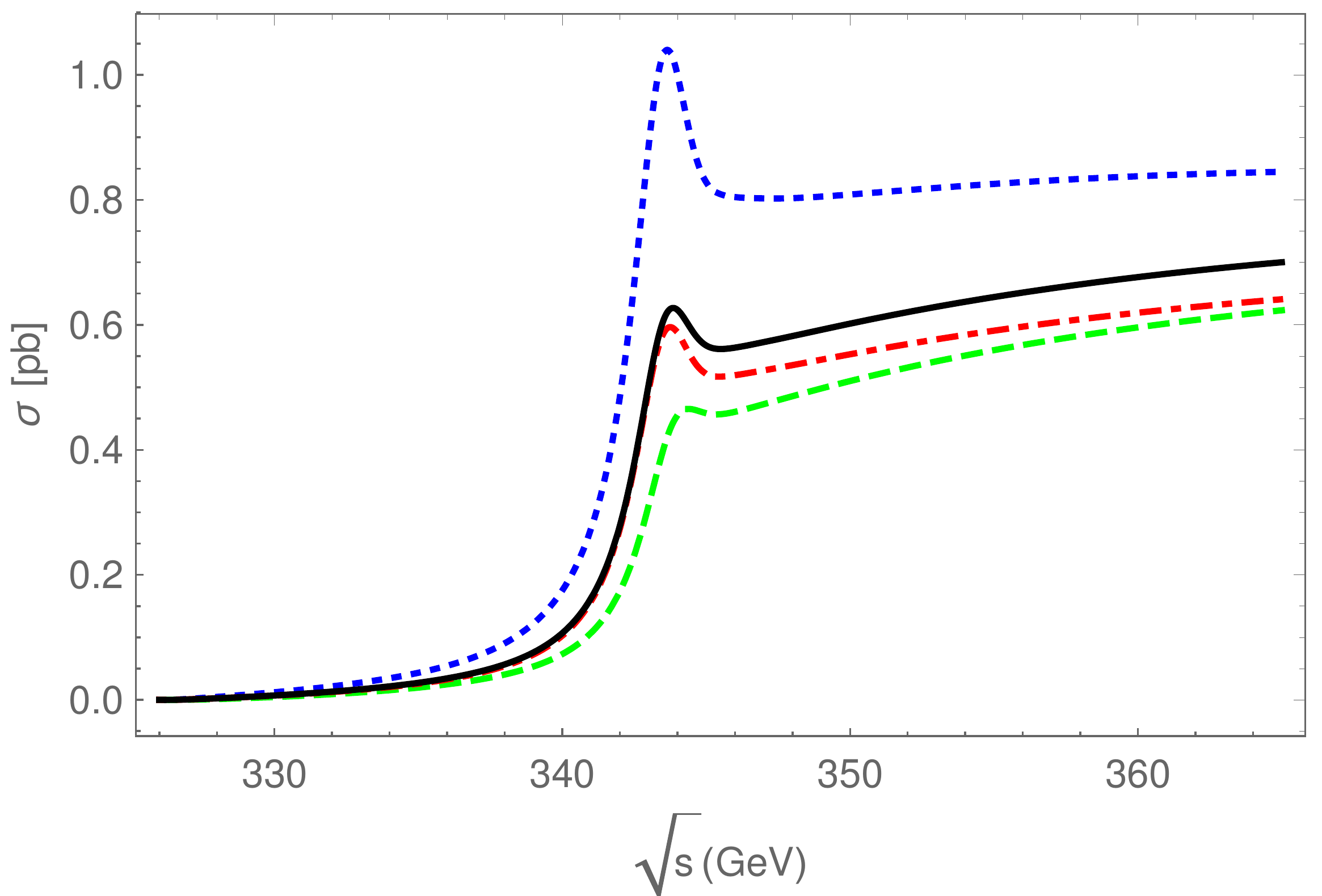}
  \caption{The QED  ISR corrections to $e^+e^- \rightarrow t\overline{t}$ ($s$-channel photon exchange) in the threshold 
  region for a PS-mass of $m_t = 172~\GeV$.
Dotted line $O(\alpha^{0})$;
Dashed line $O(\alpha)$;
Dash-dotted line $O(\alpha^{2})$;
Full line $O(\alpha^{2})$ + soft resummation; from \cite{Blumlein:2019pqb}.
  \label{fig:TT1}}
\end{figure}

\vspace*{1mm}
\noindent
For the process $e^+e^- \rightarrow t\overline{t}$ at the production threshold
we use the code {\tt QQbar\_threshold} \cite{Beneke:2016kkb,Beneke:2017rdn,Beneke:2015kwa} where N$^3$LO QCD corrections are implemented and apply the ISR effects.
The accuracy of future colliders to measure this scattering cross section
has been estimated to be $\pm 2\%$ \cite{Seidel:2013sqa,Simon:2016pwp}.
For the top-quark mass we use the PS mass of $172~\GeV$. 
The corrections are shown in Figure~\ref{fig:TT1}, they have a significant
impact on the profile of the cross section. 
The contributions due to soft photon resummation are negligible up to
$\sqrt{s} \sim 344~\GeV$ but give an additional contribution above.
Here adding soft exponentiation can change the cross section between 2 and 8\%.
The effects of the NNLO corrections and the soft photon exponentiation
both have effects of the same order or even larger than the expected 
experimental accuracy.

\section{Conclusion} 
\label{sec:5}

\vspace*{1mm}
\noindent
We have recalculated the $\mathcal{O}(\alpha^2)$ QED initial state radiation 
radiators to the $e^+ \, e^-$ annihilation into a neutral vector boson
without approximations in intermediate steps.
For many of the radiators we find closed form solutions in the form
of iterated integrals over involved, square root valued letters.
From these expressions we derived the asymptotic expansion $m_e^2 \ll s$
and compared to the earlier calculations \cite{Berends:1987ab,Blumlein:2011mi} which do not agree in the
non-logarithmic terms.
We find agreement with Ref.~\cite{Blumlein:2011mi}.
We furthermore extended the calculation of Ref.~\cite{Berends:1987ab} by adding neglected
contributions and an exact treatment of the axial vector treatment.

Furthermore, we investigated the importance of the $O(\alpha^2)$ ISR corrections
and the differences to Ref.~\cite{Berends:1987ab} for the determination of the $Z$ peak and
the $t \, \bar{t}$ threshold.
The differences to Ref.~\cite{Berends:1987ab} can become sizable if low values of $z$ are reached, e.g.
this manifests itself as a width shift of 4 MeV in the $Z$ boson width for a cut
on the invariant mass of the muon pair of $s_0=4m_\tau^2$ which is larger than
the current experimental error.
At the $t \, \bar{t}$ threshold we see large corrections in the order of or even
larger than the envisioned experimental accuracy.
Here a more detailed study is necessary to find out whether even higher order corrections are needed.

\section*{Acknowledgments}

\vspace*{1mm}
\noindent 
This project has received funding from the European Union's Horizon 2020 research and innovation programme 
under the Marie Sk\/{l}odowska-Curie grant agreement No. 764850, SAGEX, and COST action CA16201: Unraveling
new physics at the LHC through the precision frontier and from the Austrian FWF grants P 27229 and P 31952 
in part. The diagrams have been drawn using {\tt Axodraw} \cite{Collins:2016aya}.


\end{document}